\documentclass[graybox]{svmult}

\usepackage{mathptmx}       
\usepackage{helvet}         
\usepackage{courier}        
\usepackage{type1cm}        
\usepackage{makeidx}         
\usepackage{graphicx}        
\usepackage{multicol}        
\usepackage[bottom]{footmisc}

\makeindex             

\begin{document}

\title*{Complex magnetic evolution and magnetic helicity in the 
solar atmosphere}
\titlerunning{Magnetic Evolution and Helicity in Solar Atmosphere}
\author{Alexei A. Pevtsov}
\institute{Alexei A. Pevtsov \at National Solar Observatory, PO Box 62, Sunspot, NM 88349, USA, 
\email{apevtsov@nso.edu}}
%
%
\maketitle

\abstract{Solar atmosphere is a single system unified by the presence of 
large-scale magnetic fields. Topological changes in magnetic fields that 
occur in one place may have consequences for coronal heating and eruptions 
for other, even remote locations. Coronal magnetic fields also play role
in transport of  magnetic helicity from Sun's subphotosphere/upper convection 
zone to the interplanetary space.  We discuss observational evidence pertinent 
to some aspects of the solar corona being a global interconnected system, 
i.e., large-scale coronal heating due to new flux emergence, eruption of 
chromospheric filament resulting from changes in magnetic topology  triggered 
by new flux emergence, sunspots rotation as manifestation of transport of 
helicity through the photosphere, and potential consequences of re-distribution of energy from solar luminosity to the dynamo for solar cycle variations of 
solar irradiance.}              

\section{Introduction}       
\label{intro}                     
Solar atmosphere is not simply a collection of individual features. It is a 
single system unified by the presence of large-scale magnetic fields. As 
magnetic field emerges through the photosphere into the corona, it expands 
significantly forming a canopy of relatively strong magnetic 
fields overlying field-free or weaker field areas. X-ray and EUV images 
show a "network" of loops interconnecting neighboring and distant active 
regions even across the solar equator (e.g., Tadesse et al, 2011; 
Pevtsov 2000). In some respect, at any given moment the solar corona is 
completely filled by the magnetic fields at different scales and field 
strengths. Because of $\nabla \times \bf{B} = 0$ condition, there are no 
"free" magnetic polarities: every magnetic "pole" is connected to somewhere 
else. Still, observations show that shortly after its emergence, new magnetic 
flux establishes new connections with its neighbours, which implies that 
other previously existed connections would inevitably change. Thus, a 
seemingly localized flux emergence may lead to readjusting magnetic topology 
over much larger area, potentially causing additional heating and/or 
destabilizing distant coronal flux systems. 
Due to page limitations, this article is restricted to a discussion of  
effects of (localized) change in magnetic topology on coronal heating and 
remote triggering of eruptions. In addition, we also 
discuss the nature of sunspot rotation as possible indication of transport 
of helicity from below the photosphere and present consideration of the 
role of energy diverted to operate the solar dynamo in the total solar 
irradiance variations.

\section{Enhanced coronal heating in response to a remote emergence of a
new flux}
\label{cheat}

\begin{figure}[h]
\center{\includegraphics[scale=.6]{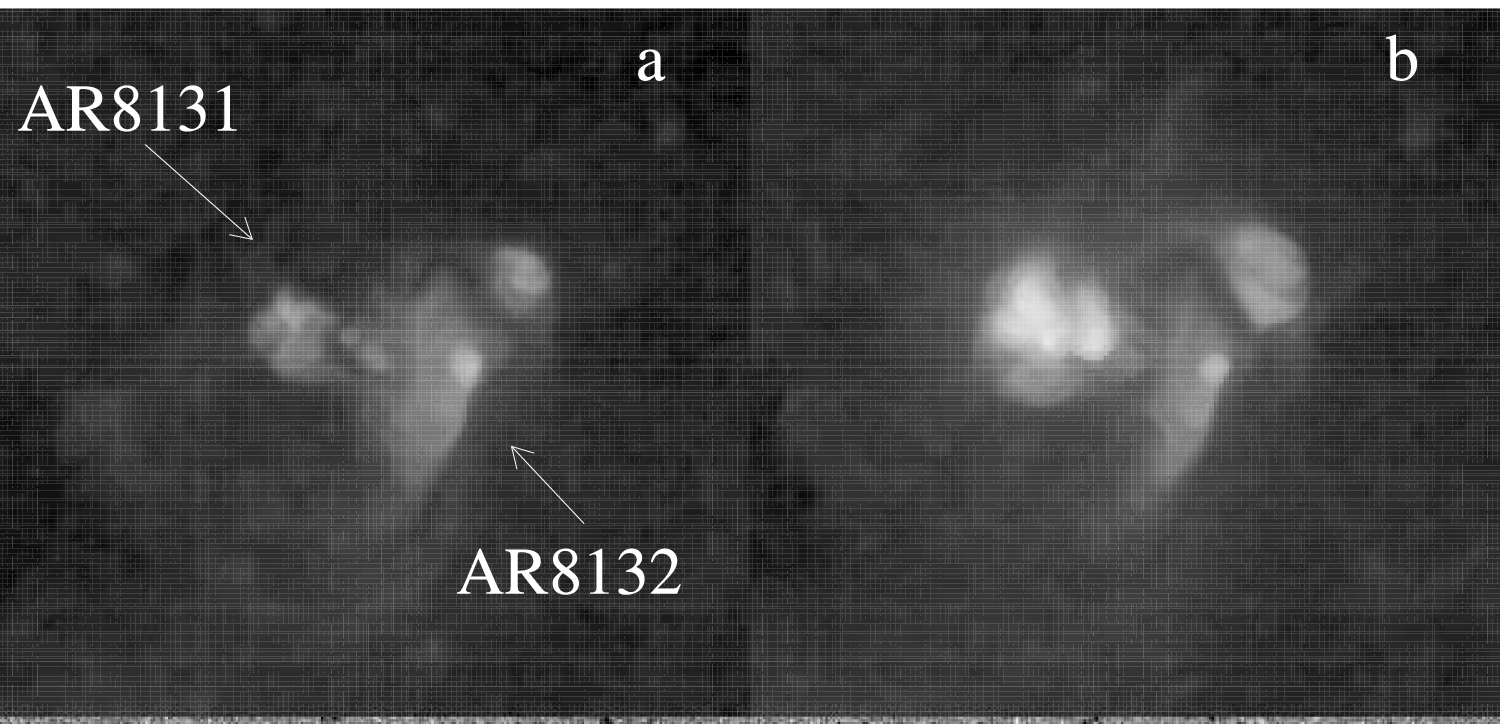}
\includegraphics[scale=.6]{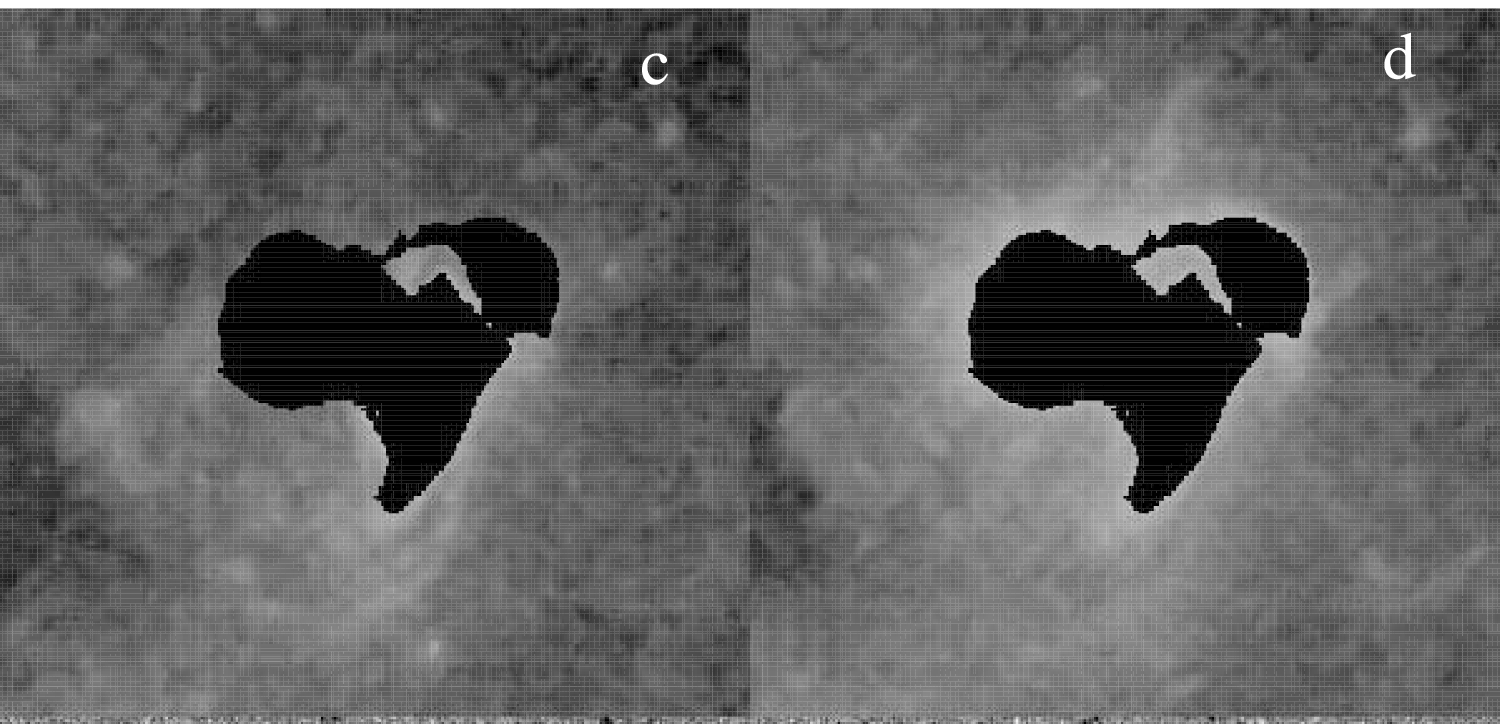}
}
\caption{X-ray images of ARs 8131 and 8132 taken by SXT on Yohkoh with Al.1 
filter on 11 January 1998, 16:08:01 UT (panels a and c), and 12 January 1998, 
00:12:55 UT (panels b and d). In lower panels, the area of brightest corona
is masked to demonstrate the increase of brightness in extended area surrounding
emerging active region. Adopted from Pevtsov and Kazachenko (2004).}
\label{fig1}       
\end{figure}

A simple look at solar images taken in EUV or X-ray wavelength bands leaves no 
doubt that magnetic fields are present almost everywhere in the corona. By its 
nature, the coronal fields maintain a constant dynamic equilibrium: changes 
in magnetic connectivity in one part of the corona, may lead to changes in 
other parts. When a new magnetic flux emerges through the photosphere, it does 
not emerge in magnetically empty corona;  its magnetic field will interact 
with pre-existing large-scale field. On smaller spatial scales, such 
interaction may transfer flux between closed and open fields, leading to 
formation of coronal jets (Moreno-Insertis et al. 2008) or coronal bright 
points 
(Kankelborg and Longcope, 1999).  If a new active region develops underneath 
a large scale magnetic field, the coronal flux system will re-adjust to 
accommodate the new flux. This re-adjustment may include development of 
new connections (for example, see Figure 6 in Longcope 2004 showing a new
loop developing between emerging active region AR9574 and existing active 
region AR9570, and Figure 4 in Pevtsov 2000 showing development of 
transequatorial loops between emerging and pre-existing active regions).  
Moore et al. (2002) observed episodic increase in brightness of 
coronal loops in the vicinity of a new flux emergence site, and have 
contributed these variations to the reconnection events associated with 
interaction between the emerging and existing magnetic flux systems. 
Pevtsov \& Acton (2001) reported increase in brightness of solar corona 
over a large fraction of solar disk, associated with the emergence of a 
single active region. Shibata et al. (1991) have suggested that the 
reconnection between emerging and pre-existing magnetic systems may result 
in heating of large-scale corona above the emerging flux. 
More realistic reconnection in 3D geometry also shows formation of area of
enhanced heating above the emerging flux (Galsgaard et al. 2005). Pevtsov 
and Kazachenko (2004) studied the emerging active region AR 8131 and its 
interaction with the existing region AR8132. They found a significant 
increase in brightness of solar corona in areas adjacent to the emerging 
flux even though there was no corresponding change in magnetic 
flux in the same area in the photosphere. Thus, for example, Yohkoh soft X-ray 
images showed about 485\% increase in X-ray intensity over the area 
encompassing two active regions, but excluding the emerging flux region itself
(Figure \ref{fig1}). 
The change occurred over the 8 hours time interval. Change in the photospheric 
flux over the same area and same time interval was about 8\%.
Pevtsov \& Kazachenko (2004) have estimated the total amount of thermal energy 
deposited in the corona as the result of interaction between the emerging and 
existing flux systems (but excluding the coronal loops directly associated 
with the emerging flux) as 2.4-3.3 $\times$ 10$^{30}$ erg. The rate of 
total thermal energy was found to be nearly constant during the early 
stages of emergence of the active region, which suggests a continuous heating.

\begin{figure}[h]
\center{\includegraphics[scale=.22]{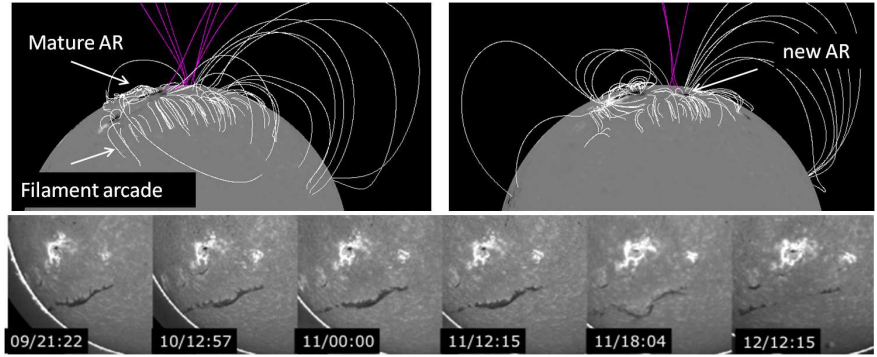}}
\caption{Magnetic field lines of filament arcade, mature active region 
(AR10830), and emerging region AR10831 from PFSS model (upper panel). Lower 
panel shows quiescent filament on different stages of its evolution on June 
9-11, 2003. Rapid rise of central part of filament prior to its eruption
can be seen on  a panel corresponding to June 11 at 18:04 UT.
Magnetic field of emerging AR reconnects with mature AR, which in its turn, 
"steals" field lines from magnetic arcade above the filament.}      
\label{fig2}       
\end{figure}

\section{Changes in large-scale magnetic connectivity and eruption of a 
filament from a distant location}

New flux emergence is often considered as a potential trigger for coronal
mass ejections (CMEs), filament eruptions, and flares (e.g., Chen and Shibata 
2000). Past statistical studies found that as much as two-thirds (Bruzek 
1952) to three-fourths (Feynman and Martin 1995) of quiescent filaments were 
de-stabilized by the birth of a nearby active region. Wang and Sheeley (1999) 
used potential field solar surface (PFSS) model to demonstrate that a 
newly-emerged magnetic flux may reconnect with the magnetic fields of arcades 
overlying chromospheric filament.  Weakening the arcade may destabilize 
the filament and lead to its eruption. 

However, the connectivity change may be indirect and more complex. 
Balasubramaniam et al. (2011) have presented case when a filament eruption was
the result of multi-step reconnection. First, a newly-formed active 
region 10381 had developed new magnetic connectivity with existing region 10380.
This new connectivity disrupted the previously existing connections between
two fluxes of opposite polarity comprising AR 10380, which, in its turn,
led to
establishing new magnetic connections between AR 10830 and 
neighbouring magnetic fluxes. The latter re-configuration weakened the 
magnetic arcade above the filament channel next to AR  
10830 and resulted in a filament eruption. Figure \ref{fig2} shows overall 
magnetic topology (based on PFSS model extrapolated from SOHO/MDI magnetograms)
and evolution of filament as observed by the ISOON H$\alpha$ 
telescope (Balasubramaniam and Pevtsov 2011).     


\section{Sunspot Rotational Motions as Indication of Helicity Transport}

Sunspot rotational motions, when a sunspot exhibits a clockwise/counter-
clockwise (CW/CCW) rotation relative to its 
geometric center, have been first reported more than a century ago 
(Kempf, 1910). Later studies by several researchers (e.g., Gnevysheva, 1941; 
Miller, 1971; Gopasyuk, 1981; Kucera, 1982; Solov'ev, 1984; Pevtsov \& 
Sattarov, 1985; Nagovitsyna \& Nagovitsyn, 1986) established typical 
properties of 
sunspot rotation including their average angular rotation rate (17$\pm$15 deg 
day$^{-1}$, e.g., Pevtsov \& Sattarov, 1985). In bipolar active regions, 
sunspots of leading and following polarity were observed to rotate in phase
either in the same or opposite direction (see Figure 1 in Pevtsov \& 
Sattarov 1985). Some sunspots exhibited change in 
the direction of their rotation, which was doubted "torsional oscillations of 
sunspots" (e.g. Gopasyuk, 1981; Solov'ev, 1984; Pevtsov \& Sattarov, 1985). 
The periods of the torsional oscillations were found to be of on order of a 
few days, although much shorter periods (of a few hours) had also been 
reported (e.g., Druzhinin et al., 1993). Some sunspots exhibited torsional 
oscillations with decreasing or increasing amplitude (e.g., Kucera, 1982; 
Pevtsov \& Sattarov, 1985). Amplitude of sunspot torsional oscillations was 
found to show a solar cycle dependency (Khutsishvili et al., 1998).
Torsional oscillations of sunspots were used to estimate the depth to which
sunspot rotational motions penetrate below the photosphere (10,000 km -- Solov'ev, 1984; and 7,500 km -- Pevtsov \& Sattarov, 1985, accordingly).  
Renewed interest in sunspot rotational motions came with high-cadence data 
from TRACE (e.g., Brown et al 2003).

It has been suggested that sunspot rotational motions may be an indication of 
helicity (twist) transport across the photosphere. According to one scenario, 
sunspot rotation may "pump" helicity to the corona leading to flares and CMEs. 
Numerical estimates indicate that the amount of helicity transported by a 
typical rotating sunspot is in agreement with the amount of helicity ejected 
by CME (e.g., Tian \& Alexander 2006). Kinetic energy of sunspot rotation is 
about 10$^{31}$ erg (Pevtsov \& Sattarov, 1985), which is comparable to 
energy of a typical flare.

Alternatively, one can hypothesize 
that sunspot rotation is a response to a removal of magnetic twist (helicity)
from the corona by flare or CME (e.g., Pevtsov 2008). In this latter 
scenario, 
subphotospheric portion of magnetic flux tube serves as a 
reservoir of helicity for the coronal portion. Prior to eruption, both parts 
are in equilibrium, but removing helicity from the corona disturbs the 
equilibrium and causes helicity to be transported from below the photosphere 
until a new equilibrium is established. Such evolution of twist (helicity)
is observed in emerging active regions (e.g., Pevtsov et al., 2003). 
Active regions with strong 
kinetic helicity below the surface are found to be more flare productive
(Reinard et al., 2010). 

\begin{figure}[h]
\center{
\vbox{
\includegraphics[scale=.22]{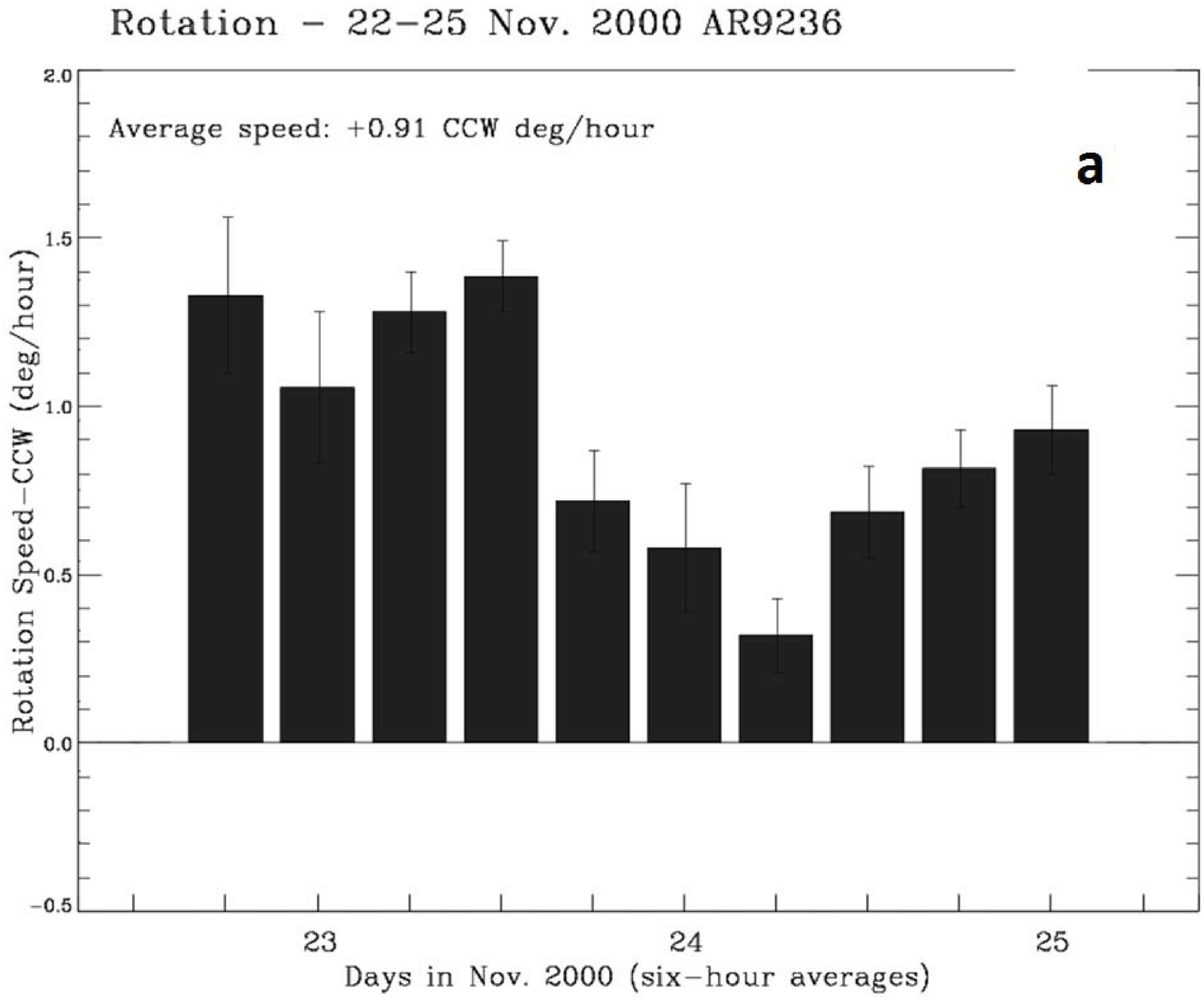}
\includegraphics[scale=.42]{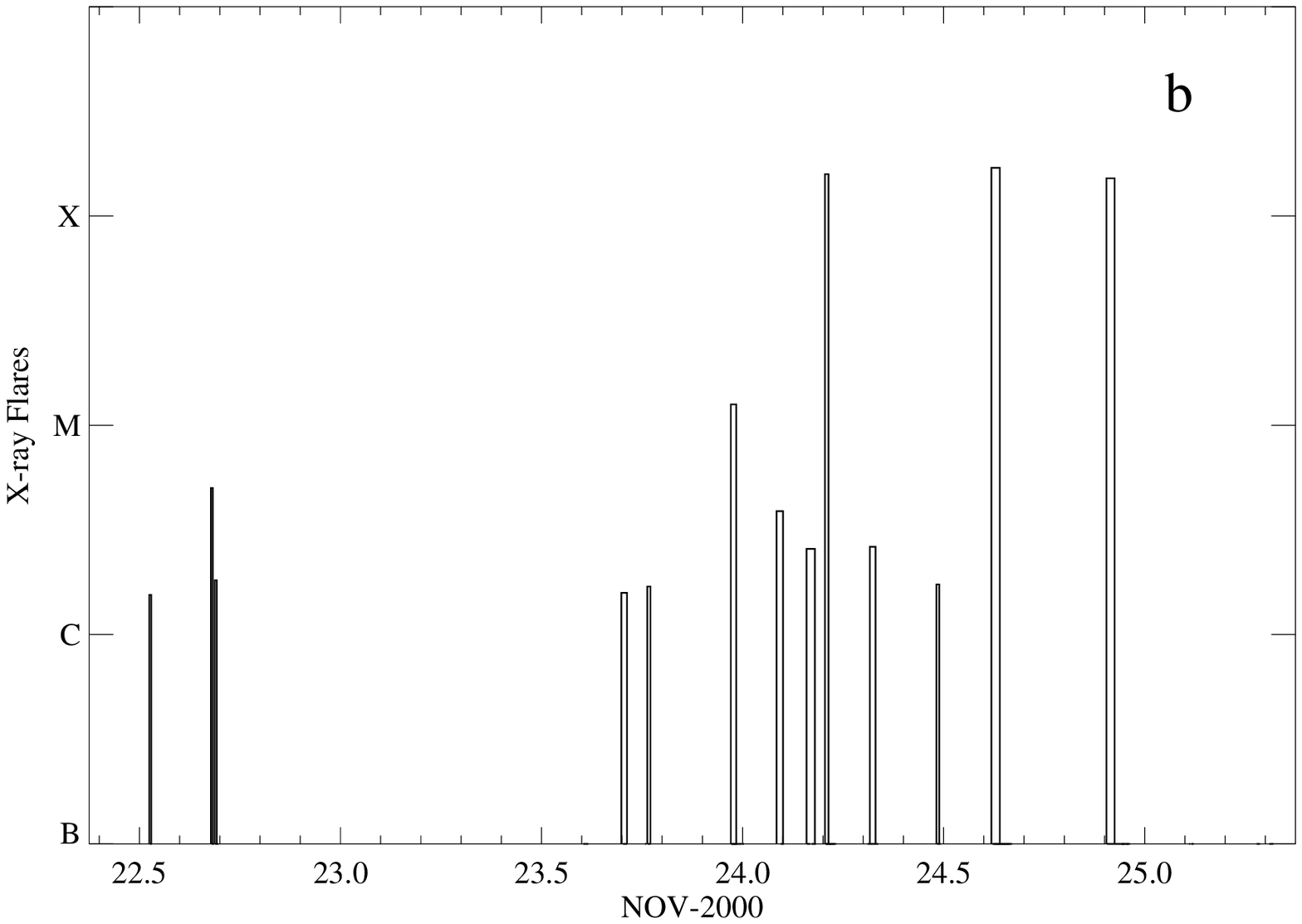}}}
\caption{Rate of sunspot rotation (a) and flare activity (b) of active region 
NOAA AR9236. Maximum X-ray flare flux and total duration of flares are shown 
on panel (b). Panel (a) is courtesy of R. Nightingale.} 
\label{fig3}       
\end{figure}

These two scenarios can be distinguished
by the timing of sunspot rotation and flare/CME eruption. If the rotating 
sunspot twists the coronal magnetic field, the flares should occur at/near the 
maximum of twist (i.e., when the sunspot rotation is strong). If the rotation 
starts after a flare/CME eruption, this might indicate that it is a response 
to helicity removal from the corona (our second scenario).
Figure \ref{fig3} shows that a period with several large flares in 
NOAA AR9236 is followed by increase in amplitude of sunspot rotation in this 
active region. This seems to be in agreement with our hypothesis that sunspot 
rotation is a response of magnetic field on helicity removal from the corona.
However, to verify the commonality of such scenario requires study of 
additional cases of sunspot rotation. It is worth noticing that the direction 
of sunspot rotation maybe hemisphere dependent. In a recent study, 
R. Nightingale
(private communication) had found that about 70\% of rotating sunspots show 
counter-clockwise rotation in Northern hemisphere. For the Southern 
hemisphere the asymmetry is weaker, with about 56\% of sunspots rotating in 
clockwise 
direction. About 15\% of sunspots in both hemispheres had shown change in the 
direction of rotation (earlier referred to as torsional oscillations of 
sunspots). The hemispheric preference in rotation of sunspots is in agreement
with well-known hemispheric helicity rule (Pevtsov et al., 1995), which provides
an indirect support for our second scenario (sunspot rotation as transport of 
helicity).

\section{Solar Dynamo and Luminosity}
\label{dynamo}

Magnetic field on the Sun is generated by the processes collectively called 
solar dynamo. In a nutshell, motions of highly conductive plasma in presence 
of seed magnetic field creates electromotive effect that further amplifies 
magnetic field. Energy that drives these flows comes from nuclear reaction 
in the core of the Sun -- same source that powers total solar luminosity. 
Thus, this energy spent on generation of magnetic field is taken out of energy 
going to luminosity. If there is no phase-shift between the production of
the magnetic field  
in the convection zone and its emergence through the photosphere, balk of 
magnetic field should be generated at/near solar maximum. Therefore, one can 
expect a dip in solar luminosity when the dynamo operation is at its maximum
because more energy is diverted to the dynamo action. How significant is the 
effect?
Rempel (2008) has estimated that the total energy of magnetic fields (E$_m$) 
stored at the base of the convection zone over 10 year solar cycle is about 
E$_m \approx$ 10$^{38}$ -- 10$^{39}$ erg. In comparison, total thermal
energy emitted by Sun over same period is 3.9$\times$10$^{33}$ erg 
$\cdot$ s$ ^{-1} \times$ 10 years $\approx$ 10$^{42}$ erg or E$_L 
\approx$ 10$^{41}$ erg per year.

Assuming that during solar maximum dynamo produces 10 times more magnetic 
field as compared with solar minimum, one can arrive to two estimates of 
magnetic energy produced in solar minimum and maximum: E$_M$(minimum) = 1.5 
$\times$ 10$^{37} $ erg and E$_M$(maximum) = 1.5 $\times$ 10$^{38}$ erg. 
By comparison with total radiative energy of the Sun, magnetic energy is only 
about 0.03\% in solar minimum, and it reaches 0.15\% in solar maximum. 
Although the magnetic energy makes such a small fraction of radiative energy, 
it is comparable in amplitude with cycle variation of total solar irradiance
and may need to be taken into consideration.   

Of course, this decrease in solar luminosity due to dynamo action is in 
anti-phase with cycle variation of TSI. However, if this expected decrease 
in luminosity is real, potentially it may offset even larger variations in 
TSI than have been observed.

\begin{acknowledgement}
National Solar Observatory (NSO) is operated by the Association of Universities
for Research in Astronomy, AURA Inc. under cooperative agreement with the
National Science Foundation.
\end{acknowledgement}
%
%
%

\end{document}